\documentclass[conference, onecolumn]{IEEEtran}
\IEEEoverridecommandlockouts
\usepackage{textcomp} 
\usepackage{xcolor}
\usepackage{amsfonts}
\usepackage{cmap}
\usepackage{titlesec}
\usepackage{amsmath,amssymb, amsthm}
\usepackage{bbm}
\usepackage{graphicx}
\usepackage{caption}
\usepackage{subcaption}
\usepackage[perpage]{footmisc}
\usepackage{cite}
\usepackage{hyperref}
\usepackage{bibleref}
\usepackage{multicol}
\usepackage[ruled, lined, longend, linesnumbered]{algorithm2e}
\usepackage{algorithmic}
\usepackage{ifthen}
\usepackage{array}

\usepackage{geometry}
\title{\textcolor{black}{\huge Covering Codes as Near-Optimal Quantizers for Distributed Hypothesis Testing Against Independence}}

\author{%
	\IEEEauthorblockN{Fatemeh~Khaledian\IEEEauthorrefmark{1},
		Reza~Asvadi\IEEEauthorrefmark{1},
		Elsa~Dupraz\IEEEauthorrefmark{2},
		and Tad~Matsumoto\IEEEauthorrefmark{2}\IEEEauthorrefmark{3}}\vspace*{5pt}
	\IEEEauthorblockA{\IEEEauthorrefmark{1}%
		Faculty of Electrical Engineering,
		Shahid Beheshti University, Tehran, Iran
	}\vspace*{1.5pt}
	\IEEEauthorblockA{\IEEEauthorrefmark{2}%
		IMT Atlantique,
		CNRS UMR 6285, Lab-STICC, Brest, France
	}\vspace*{1.5pt}
	
	\IEEEauthorblockA{\IEEEauthorrefmark{3}%
		JAIST and University of Oulu (Emeritus)
	}
}
\makeatletter
\newtheoremstyle{citestyle}%
{3pt}% space above
{3pt}% space below
{\normalfont}% body font
{}% indent amount
{\normalfont}% theorem head font
{:}% punctuation after theorem head
{.5em}% space after theorem head
{\thmname{#1}\thmnumber{ #2} \thmnote{(#3)}}% }% theorem head spec
\makeatother

\makeatletter
\newtheoremstyle{remarkstyle}%
{3pt}% space above
{3pt}% space below
{\normalfont}% body font
{}% indent amount
{\normalfont}% theorem head font
{:}% punctuation after theorem head
{.5em}% space after theorem head
{\thmname{#1}\thmnumber{ #2}}% theorem head spec
\makeatother

\theoremstyle{citestyle}
\newtheorem{thmcite}{Theorem}
\newtheorem{propcite}{Proposition}                                                                                              
\newtheorem{corcite}{Corollary}
\newtheorem{defcite}{Definition}
\newtheorem{lemmcite}{Lemma}
\theoremstyle{remarkstyle}
\newtheorem{remcite}{Remark}
\newtheorem{lemmstyle}{Lemma}
\newtheorem{propstyle}{Proposition}
\newtheorem{corstyle}{Corollary}

\def\BibTeX{{\rm B\kern-.05em{\sc i\kern-.025em b}\kern-.08em
	T\kern-.1667em\lower.7ex\hbox{E}\kern-.125emX}}
\begin{document}
\maketitle
\begin{abstract}
	We explore the problem of distributed Hypothesis Testing (DHT) against independence, focusing specifically on Binary Symmetric Sources (BSS). Our investigation aims to characterize the optimal quantizer among binary linear codes, with the objective of identifying optimal error probabilities under the Neyman-Pearson (NP) criterion for short code-length regime. We define optimality as the direct minimization of analytical expressions of error probabilities using an alternating optimization (AO) algorithm. Additionally, we provide lower and upper bounds on error probabilities, leading to the derivation of error exponents applicable to large code-length regime. \textcolor{black}{Numerical results are presented to demonstrate that, with the proposed algorithm, binary linear codes with an optimal covering radius perform near-optimally for the independence test in DHT}.
\end{abstract}
\section{Introduction}
\textcolor{black}{In} collaborative decision-making scenarios, multiple agents or sensors \textcolor{black}{send their local observations to a decision center that aims to infer a global state}. % It plays a crucial role in communication systems, optimizing resource allocation, and enhancing system reliability in dynamic environments.
\textcolor{black}{Distributed Hypothesis Testing (DHT) refers to a particular case where} the decision center's objective is to \textcolor{black}{decide} between two hypotheses, $\mathcal{H}_0$ and $\mathcal{H}_1$, leveraging \textcolor{black}{coded} version of \textcolor{black}{a} source $X$ and \textcolor{black}{a} side information $Y$. \textcolor{black}{The decision process is characterized by} Type-I and Type-II error probabilities, represented by $ \alpha_n $ and $ \beta_n $, respectively. \textcolor{black}{Information-theoretic DHT analyses} often investigate the achievable error exponent of Type-II error probability,
while \textcolor{black}{satisfying} a certain constraint $\epsilon$ on Type-I error probability.

Berger \cite{berger1979decentralized} introduced the DHT problem, \textcolor{black}{which has sparked} significant interest. \textcolor{black}{Subsequently} Ahlswede and Csiszar \cite{ahlswede1986hypothesis}, and later Han \cite{han1987hypothesis}, considered a two-node system \textcolor{black}{represented in Fig. 1}, where the first node collects data and communicates with a second node acting as the decision center over a noiseless, rate-limited link. \textcolor{black}{In \cite{ahlswede1986hypothesis}, \textcolor{black}{an achievable optimal error exponent of $\beta_n$ is \textcolor{black}{provided} for a special case of ``testing against independence"}, which is of great interest in scenarios where detecting absence of presence of statistical dependence between sources may influence further analysis or decoding}. \textcolor{black}{Several other variations of DHT problems have been explored in \cite{wigger2016testing, rahmanoptimality, xu2021distributed, salehkalaibar2020distributed, espinosa2019new, salehkalaibar2021distributed, salehkalaibardistributed}}. %investigated include those involving single sensors and multiple decision centers \cite{wigger2016testing}, multiple sensors \cite{rahmanoptimality}, binary sources \cite{haim2016binary}, general sources \cite{adamou2023information}, constant-bit communication constraints \cite{xu2021distributed}, variable-length coding \cite{salehkalaibar2020distributed}, rate-limited constraints \cite{espinosa2019new}, zero-rate compression \cite{salehkalaibar2021distributed}, or unequal-error protection codes \cite{salehkalaibardistributed}.\\

\textcolor{black}{However}, practical coding perspective of DHT has received \textcolor{black}{much less} attention compared to \textcolor{black}{the information-theoretic} aspects of the problem.
While certain studies, such as those by Haim and Kochman in \cite{haim2016binary}, have integrated linear codes into \textcolor{black}{their analysis of the error exponent}, the optimal \textcolor{black}{properties} of such codes remain uncertain. Regarding the design of practical quantizers for DHT, an iterative alternating optimization (AO) algorithm \cite{bertsekas2015parallel} is designed, considering a criterion of distributional distance (Bhattacharyya distance), to find optimal quantizers in a multiple sensor setup, as discussed in \cite{longo1990quantization}. Furthermore, \cite{chenquantization} studied the optimal scalar quantization scheme for the distributed independence test in the case of Gaussian sources. Additionally, in \cite{inan2022fundamental}, memoryless quantization methods are employed due to their memory-efficiency and low-latency characteristics \textcolor{black}{for use in a multiple-node setup}.%This \textcolor{black}{gap} \textcolor{black}{has motivated} our investigation into the two-node testing against independence setup, as exemplified by the system shown in Fig. 1. %from a perspective aimed at finding optimal linear codes.\\
\begin{figure}[t]
	\centering
	\includegraphics[scale=0.9]{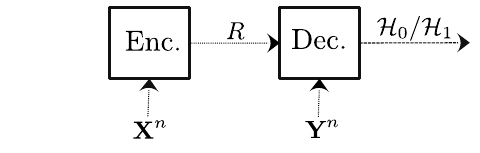}
	\captionsetup{font=footnotesize, labelfont=bf}
	\caption{Two-node system for distributed hypothesis testing problem.}	
	\label{Fig0}
\end{figure}
%According to \cite{ahlswede1986hypothesis}, a local quantization \textcolor{black}{is performed} at the first node and a global decision \textcolor{black}{which takes into account the side information $Y$} can reach the optimum Type-II error probability for DHT against independence.
%The study explores the trade-off between communication constraints and error exponents in DHT within sensor networks, culminating in lattice quantization at high communication rates.\\ 
%\textcolor{black}{Design} of binary quantizers by considering channel characteristics and using joint source-channel code (JSCC) methods is investigated in \cite{chen2004channel}.

\textcolor{black}{Despite the efforts in optimizing local quantizers in DHT \textcolor{black}{problems}, the scenario of testing against independence has garnered comparatively less attention}. This paper presents two main contributions. \textcolor{black}{First}, \textcolor{black}{it explores the utilization of linear block codes as the local quantizer component, \textcolor{black}{by relying on} exact analytical expressions for Type-I and Type-II error probabilities provided in \textcolor{black}{\cite{Elsa2023}}}. \textcolor{black}{\textcolor{black}{An iterative AO algorithm is proposed to identify} the optimal characteristics of the binary local quantizer by optimizing the \textit{coset leader spectrum} of linear block codes while also optimizing a decision rule under Neyman-Pearson (NP) criterion}. %This process leads to finding the minimum value of Type-II error probability through the optimized threshold and also the minimum possible value of Type-I error probability using the optimized coset leaders spectrum of the linear codes with respect to a certain constraint $\epsilon$.
\textcolor{black}{Then, it derives error exponents for Type-I and Type-II error probabilities using binary linear codes as the local quantizer component. In \textcolor{black}{addition} to the exact error exponent derivations, to grasp a general tendency  of Type-I and Type-II error exponents, upper and lower bounds of Type-I and Type-II error probabilities \textcolor{black}{are derived.}}
%The rest of the paper is organized as follows. In Section \ref{section 2}, the problem definition, along with a brief description of the properties of linear codes required in the following sections, is presented. In Section \ref{section 3}, an AO algorithm \textcolor{black}{is introduced} to identify optimal quantizer under the \textcolor{black}{NP} criterion. Section \ref{section 4} \textcolor{black}{presents} numerical examples. Finally, Section \ref{section 5} concludes the paper with a summary of key findings and concluding statements.
\section{Problem Statement}\label{section 2}
\subsection{DHT against independence for \textcolor{black}{Binary Symmetric Sources}}
Consider the system depicted in Fig. \ref{Fig0}, where the first and second nodes observe random vectors $\textbf{X}^n$ and $\textbf{Y}^n$ of length $n$, respectively. % Each component of these vectors takes values from \textcolor{black}{a binary alphabet given by $\left\{0,1\right\}$}. 
%Random samples $(X_i, Y_i)$ are independently distributed across \textcolor{black}{the time index} $i$, according to one of two possible joint distributions
%\begin{align}
%	\mathcal{H}_0&: \textcolor{black}{\left(X,Y\right)} \sim P_{XY}(x,y), \\ 
%	\mathcal{H}_1&: \textcolor{black}{\left(X,Y\right)} \sim  \textcolor{black}{P_{X}(x)P_{Y}(y)}
%\end{align}
%\textcolor{black}{for each $i \in \mathbb{N}$}. \textcolor{black}{Hypothesis} $ \mathcal{H}_1 $ represents the case of independence, where the related distribution is equal to the product of marginal distributions \textcolor{black}{under hypothesis}  \textcolor{black}{$\mathcal{H}_0$}. \textcolor{black}{This scenario is referred to as testing against independence}.
\textcolor{black}{As a Binary Symmetric Source (BSS) case,} consider the scenario \textcolor{black}{X and Y take values in a binary alphabet, and} $P_X(x) = P_Y(y) \sim \text{Bern}(\frac{1}{2})$. Although the marginal distributions of $X$ and $Y$ are identical, $Y$ may represent a noisy version of $X$. Denote $ Y \triangleq X \oplus W $, \textcolor{black}{where} $ \oplus  $ \textcolor{black}{is the} \textcolor{black}{summation} over \textcolor{black}{the} binary field. \textcolor{black}{Consider the following two hypotheses,}
\begin{align}
	\mathcal{H}_0&: W \sim \text{Bern}\left(p_0\right)\!,\\
	\mathcal{H}_1&: W \sim \text{Bern}\left(p_1\right)\!,
\end{align}
\textcolor{black}{where $\text{Bern}(p)$ denotes the Bernoulli distribution with the parameter $p$, and $0 \leq p \leq 1$}. \textcolor{black}{We also} assume throughout \textcolor{black}{the} paper $ 0 \leq p_0 <  \frac{1}{2} $. \textcolor{black}{Obviously} \textcolor{black}{$\mathcal{H}_1$ indicates that $X$ and $Y$ are \textcolor{black}{independent} when $p_1=\frac{1}{2}$}.

As illustrated in Fig. \ref{Fig0}, the first node \textcolor{black}{transmits a coded version of} $\mathbf{X}^n$ under a communication rate $R$ over a noiseless channel. The second node makes a decision based on the \textcolor{black}{coded} version of $\mathbf{X}^n$ and \textcolor{black}{the side information vector} $\mathbf{Y}^n$. This decision may result in two types of errors, with their probabilities defined as follows
\begin{align}
	\alpha_n &= Pr(\mathcal{H}_1 | \mathcal{H}_0),\\
	\beta_n  &= Pr(\mathcal{H}_0 | \mathcal{H}_1). 
\end{align}
\textcolor{black}{In these expressions,} $\alpha_n$ \textcolor{black}{is} Type-I error probability and $\beta_n$ is Type-II error probability. \textcolor{black}{In \cite{ahlswede1986hypothesis}, it is demonstrated that for testing against independence, information-theoretic quantizer-based schemes are optimal in minimizing Type-II error probability $\beta_n$ under the constraint $\alpha_n \leq \epsilon$.}
\subsection{Linear \textcolor{black}{block} codes and useful properties}
%According to \cite{ahlswede1986hypothesis}, \textcolor{black}{identifying} the optimal \textcolor{black}{error exponent} necessitates an optimal local quantizer. Exploiting the robust algebraic structures inherent in linear codes allows for the effective utilization of mathematical tools and techniques for the analysis and the design \textcolor{black}{of the quantizer}. \textcolor{black}{Hence}, \textcolor{black}{we restrict our investigations within the} linear codes as the quantizer component.\\
Consider a binary linear code $\mathcal{C}$ defined by a $k \times n$ \textcolor{black}{generator} matrix $\mathbf{G}$ with a rate of $R=\frac{k}{n}$ \textcolor{black}{as the binary quantizer  component\cite{richardson2008modern}}. According to the standard array concept \cite{lin2004error}, the minimum Hamming weight vector \textcolor{black}{$d_H\left(\cdot\right)$} in each \textit{coset} \textcolor{black}{$ \mathcal{C}_\mathbf{s} $} associated to the syndrome $\mathbf{s}$ is referred to as the \textit{coset leader} defined as
\begin{equation}
	L(\textcolor{black}{\mathcal{C}_\mathbf{s}}) \triangleq \arg \min_{\substack{\mathbf{z} \in \textcolor{black}{\mathcal{C}_\mathbf{s}}}} d_H\left(\mathbf{z}\right)\!,\nonumber
\end{equation}
and all coset leaders are denoted by \textcolor{black}{$\mathcal{L} = \left\{L\left(\mathcal{C}_\mathbf{s}\right): \text{for all possible $\mathbf{s}$}  \right\}$}.
\begin{defcite}[\textit{\textcolor{black}{covering radius}} \cite{cohen1997covering}]
	The covering radius of a code $\textcolor{black}{\mathcal{C} \subseteq \left\{0,1\right\}^n}$ is the smallest integer such that every vector $\mathbf{x} \in \left\{0,1\right\}^n$ is covered by a radius $\rho$ Hamming ball centered at a point $\mathbf{c} \in \mathcal{C}$, i.e.,
	\begin{equation}
		\rho(\mathcal{C}) = \max_{\mathbf{x} \in \{0,1\}^n} \min_{\mathbf{c} \in \mathcal{C}} d_H(\mathbf{x},\mathbf{c}).\nonumber
	\end{equation}
	%where $ d_H(\cdot)) $ denotes the Hamming distance.
\end{defcite}
\begin{defcite}[\textit{\textcolor{black}{coset leader spectrum}}]
	\textcolor{black}{Let $\mathcal{C}$ be a binary linear code. Its \textit{coset leader spectrum}  $\textcolor{black}{\mathbf{N}=\left(N_0,N_1,..., N_{\rho}\right)}$ is a vector of length $\rho + 1 $, where
		\begin{equation}
			\textcolor{black}{N}_i = |\left\{L \in \mathcal{L} : d_H\left(L\right)=i\right\}|,\nonumber
		\end{equation}
		i.e., the $i$-th component of \textcolor{black}{$\mathbf{N}$} is the number of coset leaders of weight $i$ in $\mathcal{L}$. }
\end{defcite}
We represent a binary linear code associated with the generator matrix $\mathbf{G}$ and the covering radius $\rho$ as $[n, k]_\rho$ throughout the paper.
\iffalse
\begin{defcite}[\textit{sphere-covering bound}\cite{cohen1985covering}]
	If an $ [n,k]_\rho $ code exists,
	\begin{equation}
		\sum_{i=0}^{\rho}{n \choose i} \geq 2^{(n-k)}.\label{def3}
	\end{equation} 
\end{defcite}
\begin{defcite}[\textit{singleton Bound}\cite{cohen1985covering}]
	If an $ [n,k]_\rho $ code exists, then $ \rho \leq n-k $.\label{def4}
\end{defcite}
\fi
\textcolor{black}{ For an $[n,k]_\rho$, It is shown that $\sum_{i=0}^{\rho}{n \choose i} \geq 2^{(n-k)}$ called sphere-covering bound \cite{cohen1997covering}, and $ \rho \leq n-k $ called Singleton bound \cite{cohen1997covering}. These two inequalities} provide general lower and upper bounds for the covering radius $\rho$, respectively. 
%\begin{prop}
%	The optimal error exponent of the DHT against independence problem in the binary symmetric sources model and for any rate $ R \geq 0 $ is obtained as follows, 
%	\begin{equation}
	%		\theta^*\left(R\right)=1-H_b\left(H_b^{-1}\left(1-R\right)*\alpha\right).
	%	\end{equation}
%Achievability proof: The optimal error exponent of the DHT against independence problem is evaluated in [],
%\begin{equation}
%	\theta^*\left(R\right)=\max_{\substack{P_{U|X}:\\ |\mathcal{U} \leq |\mathbb{X}+1 \\ R \geq I\left(U;X\right)}} I\left(U;Y\right).
%\end{equation}
%According to [], we choose an auxiliary random variable $ U $ as the output of a binary symmetric channel (BSC) with cross-over probability $ \alpha $ when the input is $  X $, 
%\begin{equation}
%  U=X+D \; \; ; \; D \sim \text{Bern}\left(d\right). \nonumber
%\end{equation}
%If we consider the BSS model,
%\begin{equation}
%Y=X+W \; \; ; \; W \sim \text{Bern}\left(\alpha\right)
%\end{equation}
%since, $ U $ and $ Y $ can be taught to be connected through a binary symmetric channel with cross-over probability $ \alpha * d $ , Hence we can write:
%\begin{align}
%	\theta^*\left(R\right)&=\max \left(1-H_b\left(U|Y\right)\right) \\ 
%	                      &=1-H_b\left(\alpha * d\right)  \label{8}\\
%	                      &=1-H_b\left(H_b^{-1}\left(1-R\right)*d\right). \label{9}	                      
%\end{align}
%Equality of $\left(\ref{8}\right)$ comes from the assumption of the BSS model, however $\left(\ref{9}\right)$ comes from choosing a BSC as the optimal test channel in the binary case.
%\end{prop}
\section{Analysis of Optimum Quantizer}\label{section 3}
\subsection{The quantizer structure under Neyman-Pearson criterion}
%Design of an optimal quantizer for DHT against independence involves \textcolor{black}{the analysis} on the local quantization process at Node A, specifically operating on its observations \cite{ahlswede1986hypothesis}. \textcolor{black}{In this paper}, the characteristics of the quantizer are chosen based on the global observation model to enhance the overall system performance \cite{longo1990quantization}.\\
%It is also worth mentioning this point that a critical conclusion of Shannon's rate-distortion theory is that better performance is always achievable by coding vectors instead of scalars, even if the data source is memoryless. A vector quantization system aims to produce the best reproduction vector for a given rate $R$.
\textcolor{black}{The optimal binary quantizer for minimizing the average distortion criterion is known as Minimum Distance (MD) quantizer \cite{gray1984vector}. An MD quantizer implemented by a binary linear block code operates so that for a source vector \( \mathbf{x}^n \), the quantized vector \( \mathbf{u}^k_q \) is given by
	\begin{equation}
		\mathbf{u}^k_q = \arg\min_{\mathbf{u}^k \in \textcolor{black}{\left\{0,1\right\}^k}}  d_H\left(\mathbf{x}^n, \mathbf{x}_q^n\right)\!,\nonumber
	\end{equation}
	where \( \mathbf{x}_q^n = \mathbf{u}^k_{\textcolor{black}{q}}\mathbf{G} \)}. % and $ \mathbf{u}^k_q $ denotes the selected quantized vector.
% However $ E\left(d_H\left(\cdot\right)\right) $ is the average Hamming distance, and $ \mathbf{u}^k_q $ is the chosen compressed vector for a given rate $ R=k/n $.
%Designing an effective quantizer based on the aforementioned property is analogous to creating a covering-good code, that is, the arrangement of codewords ensures that no vector is significantly distant from its nearest codeword \cite{graham1985covering}. While linear codes exhibit asymptotic covering-good properties \cite{cohen1985covering}, our focus is on exploring the existence of hypothetical binary linear codes that demonstrate favorable performance in finite sample scenarios.
In the design of the \textcolor{black}{rule to decide between hypotheses $\mathcal{H}_0$ and $\mathcal{H}_1$} various \textcolor{black}{methods} are available. \textcolor{black}{Given that the probability distribution of the \textcolor{black}{noise $W$} is known under both $\mathcal{H}_0$ and $\mathcal{H}_1$, we use the NP lemma \cite{lehmann1986testing} which minimizes $\beta_n$ under a certain constraint on $\alpha_n$}. By considering the BSS model \textcolor{black}{as defined in Section II-A}, \textcolor{black}{the NP lemma provides} following criterion \cite{haim2016binary}
\begin{equation}
	\sum_{j=1}^{n}\left(x_{q,j} \oplus y_j\right) \lessgtr \gamma_t,\nonumber
\end{equation}
\textcolor{black}{where \( \gamma_t \in \mathbb{N} \) is an integer threshold chosen to minimize \( \beta_n \) while satisfying the constraint \( \alpha_n \leq \epsilon \)}. %\textcolor{purple}{The threshold is \textcolor{black}{determined so that} Type-II error \textcolor{black}{probability is minimized} while \textcolor{black}{keeping} Type-I error probability \textcolor{black}{lower than or equal to a specified value} $\epsilon$.}
%This rule provides a principle approach to optimize the decision \textcolor{purple}{threshold $\gamma_t$}. % \textcolor{black}{in this paper}.
\textcolor{black}{Note that the symbol $<$ indicates a decision in favor of $H_0$, while $>$ indicates a decision in favor of $H_1$}.\\
\textcolor{black}{We next} consider exact analytical expressions for Type-I and Type-II error probabilities \textcolor{black}{provided} in \cite{Elsa2023} \textcolor{black}{for the MD quantizer}.
Specifically in the context of testing against independence, the expressions \textcolor{black}{of \cite{Elsa2023}} can be simplified as follows
\begin{align}
	\alpha_n &=\frac{1}{N}\sum_{\gamma=\gamma_t+1}^{n} \sum_{i=0}^{\textcolor{black}{\rho}} \sum_{u=0}^{\min\left(\gamma,i\right)} \Gamma_{\gamma,i,u}  p_0^{\textcolor{black}{j}} \left(1-p_0\right)^{n-\textcolor{black}{j}} \textcolor{black}{N_i}, \label{7}\\
	\beta_n &= \frac{1}{N} \left(\frac{1}{2}\right)^n \sum_{\gamma=0}^{\gamma_t} \sum_{i=0}^{\textcolor{black}{\rho}} \sum_{u=0}^{\min\left(\gamma,i\right)} \Gamma_{\gamma,i,u} \textcolor{black}{N_i}, \label{8}
\end{align}
where $ \Gamma_{\gamma,i,u}= {i \choose u} {n-i \choose \gamma-u} $ and $ \textcolor{black}{j}=i+\gamma-2u $. %\textcolor{black}{Further},  $N_i$ represents the number of source vector $\mathbf{x}^n$ with a weight $i$ in the decision region of the all-zero codeword. This parameter can be interpreted as the number of the \textit{coset leaders} with a weight of $i \in [0,\textcolor{black}{\rho}]$. 
\textcolor{black}{The total number of the \textit{coset leaders} \textcolor{black}{is} denoted by $N = \sum_{i=0}^{\textcolor{black}{\rho}} N_i$.} Additionally, for the weight of $ i $, $\Gamma_{\gamma,i,u}$ represents \textcolor{black}{the} number of all possible vectors of weight $ \gamma $ with $u$ 1's in common with $ \mathbf{x}^n $.
\iffalse
\begin{thmcite}[\cite{cohen1997covering}]
	The covering radius of a linear code is the largest of \textcolor{black}{the weights of the \textit{coset leaders}}.\label{theorem1}
\end{thmcite}
\begin{remcite}
	According to Theorem \ref{theorem1}, the value $d_{\text{max}}$ in equations (\ref{7}) and (\ref{8}) corresponds to the covering radius $\rho$ of a linear code.% \textcolor{black}{which further implies $N=2^{\left(n-k\right)}$ with $d_{\max}=\rho$}. %However, assuming a uniform distribution over codewords and considering the lower bound in (\ref{def3}), we can consider $N$ as $2^{(n-k)}$.
\end{remcite}
\fi
\subsection{Short code-length regime }
\textcolor{black}{In \cite{Elsa2023}, the performance of the quantizer was \textcolor{black}{evaluated for specific linear codes}, notably BCH codes, \textcolor{black}{with} specific coset leader spectrums. \textcolor{black}{Here, alternatively}, our \textcolor{black}{objective} is to formulate an optimization problem \textit{only under} \textcolor{black}{generic constraints: Minimize $\beta_n$ while $\alpha_n \leq \epsilon$}.}
%Although the resulting code is hypothetical due to unknown practical constructions, it provides a lower bound on the performance of the quantizer and allows for comparison with existing linear codes}.

\textcolor{black}{Consider} $ \mathbf{N} = (N_0, \ldots, N_{\rho}) $ as the \textit{coset leader spectrum} of \textcolor{black}{an} hypothetical binary linear code satisfying the condition $N_0 + \ldots + N_{\rho} = N$.
%\begin{equation}
%	N_0 + \ldots + N_{\rho} = N.\nonumber
%\end{equation}
In the following, (\ref{7}) and (\ref{8}) \textcolor{black}{can be rewritten as}   
\begin{align}
\alpha_n&=\frac{1}{N}\sum_{i=0}^{\textcolor{black}{\rho}}W_{\alpha_i}N_{i},\label{24}\\
\beta_n&=\frac{1}{N}\sum_{i=0}^{\textcolor{black}{\rho}}W_{\beta_i}N_{i}\label{25},
\end{align}
where
\begin{align}
W_{\alpha_i}&=\sum_{\gamma=\gamma_t+1}^{n} \sum_{u=0}^{\min\left(\gamma,i\right)} \Gamma_{\gamma,i,u} p_0^{\textcolor{black}{j}_i} \left(1-p_0\right)^{\left(n-\textcolor{black}{j}_i\right)},\label{5}\\
W_{\beta_i}&=\left(\frac{1}{2}\right)^n\sum_{\gamma=0}^{\gamma_t} \sum_{u=0}^{\min\left(\gamma,i\right)} \Gamma_{\gamma,i,u}.\label{6} 
\end{align}
\begin{lemmstyle}\textcolor{black}{
	\textcolor{black}{With the} blocklength $ n $ and an integer threshold $ 0 \textcolor{black}{\leq} \gamma_t \leq n $, Type-II error probability can be expressed as:
	\begin{equation}
		\beta_n =\left(\frac{1}{2}\right)^n\sum_{\gamma=0}^{\gamma_t}{n \choose \gamma}.\label{beta}
		\end{equation}}\end{lemmstyle}
		\vspace{-10pt}
\textit{proof}: The proof is given in Appendix A.

%It is \textcolor{black}{found} that the coset leader spectrum \textcolor{black}{does not} affect the Type-II probability; it \textcolor{black}{only} depends on the block length $ n $ and the decision threshold $ \gamma_t $. \\
%\textcolor{black}{An} alternating optimization problem  \textcolor{black}{is formulated} by incorporating the \textcolor{black}{NP} criterion \textcolor{black}{as follows}
According to \textcolor{black}{Lemma 1}, minimizing $\beta_n$ is equivalent to minimizing the decision threshold $\gamma_t$.
%while ensuring $\alpha_n \leq \epsilon$.
\textcolor{black}{First of all, suppose our goal is to satisfy the constraint $\alpha_n \leq \epsilon$ for a possible minimum threshold $\gamma_t$. In this case, we can formulate a minimization problem as follows} 
%This approach leads to an AO algorithm that achieves both objectives simultaneously. The core of the AO algorithm  can be described as follows  		
\begin{equation}
\textcolor{black}{\min_{\{\mathbf{N}_{i}\}}} \;\; e=\sum_{i=\textcolor{black}{1}}^{\textcolor{black}{\rho}} W_{\alpha_i} N_{i} \label{19},
\end{equation}
subject to,
%$ \begin{array}{l}
%	\left\{
%	\begin{align}
	%		&0 \leq N_{w_i} \leq {n \choose w_i}, \\
	%		&\sum_{w_i=1}^{d_{\text{max}}} N_{w_i} = N-1, \\
	%		&-\sum_{w_i=0}^{d_{\text{max}}} W_{\alpha_i}N_{w_i} \leq N\left(\epsilon - 1\right).
	%	\end{aligned}
%	\right.
%\end{array} $\\
\[
\qquad \qquad \left\{
\begin{aligned}
	&0\leq N_{i} \leq {n \choose i}\; ; \; 1 \leq i \leq \rho  &&\text{(\textcolor{black}{12a})} \\
	&\sum_{i=\textcolor{black}{1}}^{\textcolor{black}{\rho}} N_{i} = N-1, \hfill &&\text{(12b)}\\
	%&\sum_{i=1}^{d_{\text{max}}} W_{\alpha_i} N_{i} \leq N\epsilon, \hfill &&\text{(19c)}
\end{aligned}
\right.
\]
where $ W_{\alpha_i} $ is defined as in (\ref{5}). 
\textcolor{black}{In this formulation,} (12a) \textcolor{black}{comes from the finite number of vectors with a length of $n$ and a weight of $i$}, additionally (12b) \textcolor{black}{stems from \textcolor{black}{the fact} $N_0=1$}{\footnote{\textcolor{black}{\textcolor{black}{Particularly}, $N_0$ represents the all-zero codeword of \textcolor{black}{the} binary linear code \textcolor{black}{considered, hence $N_0=1$}.}}}.
%Second, consider another optimization problem as follows,
%\begin{align}
%\text{Minimize} &\; \; \; E_2=1-(\frac{1}{N})\sum_{w_i=0}^{d_{max}} W_{\alpha_i} N_{w_i} \label{21}\\
%\text{subject to}& \; \; \; \left\{\begin{array}{l}
	%0 \leq N_{w_i} \leq {n \choose w_i},\\ \\
	%\sum_{w_i=0}^{d_{max}} N_{w_i}= N-1,\\ \\
	%1-\frac{1}{N}\sum_{i=0}^{d_{max}} W_{\alpha_i}N_{w_i} \leq \alpha_n.\label{22}	
	%\end{array}\right.
	%\end{align}
	%where the coefficients $ W_{\alpha_i} $, $ W_{\beta_i} $ and the parameters $ \Gamma_{\gamma,w_i,u} $, $ k_i $ are mentioned in (\ref{17}), (\ref{18}), (\ref{19}) and (\ref{20}).\\
	
	By solving the minimization problem outlined in (\ref{19}), (12a) and (12b), aimed at optimizing the \textit{coset leader spectrum} $\textcolor{black}{\mathbf{N}}^*$ of an hypothetical binary linear code \textcolor{black}{$[n,k]_{\rho^*}$}, we can meet the constraint $\alpha_n \leq \epsilon$. Notably, the problem described in (\ref{19}), along with constraints (12a) and (12b), falls into the category of integer linear programming (ILP) problems.  
	\begin{corstyle} \label{remark2}
		%According to \textcolor{black}{Theorem 1} and \textcolor{black}{Corollary 1}, 
		\textcolor{black}{Given an $[n,k]_{\rho^*}$},
		there exists a polynomial-time algorithm that finds an optimal solution for the ILP problem \footnote{Generally, ILP problems are categorized as \textcolor{black}{Nondeterministic Polynomial time-Complete} \cite[Th. 18.1]{schrijver1998theory}. In practice, a combination of branch and bound with LP-relaxation methods like cutting-planes is employed.} 
		described in (\ref{19}), \textcolor{black}{with the constraints} \textcolor{black}{(12a) and (12b)}. \textcolor{black}{The solution is} a vector \textcolor{black}{with components}, at most, $ 6\left(n\textcolor{black}{-k}\right)^3\textcolor{black}{\Phi}$.\footnote{\textcolor{black}{$\Phi$ represents the facet complexity of a rational polyhedron associated with the ILP problem, defined by $P = \{x \mid Ax \leq b\}$, where it is equal to the maximum row size of the matrix $[A \;\; b]$ \cite{schrijver1998theory}}.}
		\end{corstyle}
		\textit{proof}: \textcolor{black}{The proof is directly derived from \cite[Theorem 16.2]{schrijver1998theory} and \cite[Corollary 17.1c]{schrijver1998theory}, according to the Singleton bound. \hfill $\blacksquare$}  
		
	\textcolor{black}{To determine the optimal decision threshold $\gamma_t^{\textcolor{black}{*}}$ while ensuring that the constraint $\alpha_n \leq \epsilon$ is satisfied, we propose an AO algorithm, outlined in Algorithm 1.} 
	%\begin{thmcite}[\textcolor{black}{[21, Theorem 16.2]}]
	%There is a polynomial algorithm, given a rational system $ Ax \leq b $ defining a \textcolor{black}{finite} integral polyhedron and given a rational vector $ c $, \textcolor{black}{that} finds an optimum solution for the ILP-problem as $ \max \left\{cx|Ax \leq b; x \; \text{\textcolor{black}{integer}}\right\} $. %\textcolor{black}{(if it is finite)}.
	%\end{thmcite}
	%\begin{corcite}[\textcolor{black}{[21, Corollary 17.1c.]}]
	%Let $ P $ be a rational polyhedron in $\mathbb{R}^n $ of facet complexity\footnote{Facet complexity of a polyhedron defined by $ P=\left\{x|Ax \leq b\right\} $ is equal to the maximum row size of the matrix $ [A \;\; b] $ \cite{schrijver1998theory}.} $ \phi $, and let $ c \in Q^n $. If $ \max \left\{cx| x \in P; x \; \text{\textcolor{black}{integer}}\right\} $ is finite, it is attained by a vector of size at most $ 6n^3\phi$.  
	%\end{corcite}
	\begin{algorithm}[t]
		\caption{\textcolor{black}{Coset Leader Spectrum Optimization}}\label{alg2}
		\renewcommand{\baselinestretch}{1.25}\selectfont
		\textbf{Procedure}{ Integer Linear Programming} ($n, k, p_0, \epsilon$)\\
		\textbf{Set} $\gamma_t=0$;\\
		%Compute $\textbf{W}_{\alpha}$ and initialize $\mathbf{N}_i$;\\
		Define the ILP problem in (\ref{19}), (12a) and (12b);\\
		\textbf{Solve} the ILP for $\gamma_t = 0$;\\
		Compute $\alpha_n$ from (\ref{24});\\
		
		\If{$\alpha_n \leq \epsilon$}{%
			\textbf{return} $\gamma_t^*=0$, $\mathbf{N}^*$;
		}
		\Else{%
			$\gamma_t = \gamma_t + 1$;
			
			\While{$\alpha_n > \epsilon$}{%
				Update $\textbf{W}_{\alpha}$ and initialize $\mathbf{N}_i$;\\
				\textbf{Solve} the ILP;\\
				Update $\alpha_n$;\\
				$\gamma_t = \gamma_t + 1$;
			}
		}
		\textbf{return} $\gamma_t^*$, $\mathbf{N}^*$;\\
		\textbf{end Procedure}
	\end{algorithm}
	\subsection{Large code-length regime}
	%\textcolor{black}{To identify significant parameters for the local quantizer as defined in Section III-A, we investigate the error exponents for $\alpha_n$ and $\beta_n$, denoted by $E_{0,MD}$ and $E_{1,MD}$ respectively, within a large-code length regime. The following proposition is directly derived from the analytical expressions in \cite{Elsa2023}.}
	\textcolor{black}{\textcolor{black}{Following the discussions in the previous Section for the local quantizers built from linear block codes first, we then investigate characteristics of the optimal quantizer} for a large code-length regime. %Proposition 1 can be directly derived from \cite{Elsa2023}}.
	\begin{propstyle}
		\textcolor{black}{% \textcolor{black}{There are} the following pair of error exponents $\left(E_{0,MD}\left(\gamma_t,n,p_0\right), E_{1,MD}\left(\gamma_t,n\right)\right)$ as
			%\begin{align}
			%	E_{0,MD}\left(\gamma_t,n,\textcolor{black}{p_0}\right) &= D_b\left(\frac{\gamma_t}{n}||p_0\right)\!,\label{E_0} \\
			%   E_{1,MD}\left(\gamma_t,n\right) &=1 - H_b\left(\frac{\gamma_t}{n}\right)\! \label{E_1}
			%\end{align} 
			%with $p_0 < \frac{\gamma_t}{n} < \frac{1}{2}$. \textcolor{black}{Here}, \textcolor{black}{$D_b(p||q)$ represents the binary Kullback-Leibler divergence between the probability pair $\left(p,q\right)$, where $\left(p,q\right) \in \left[0,1\right]$, and $H_b\left(\cdot\right)$ denotes the binary entropy function}.	
			Let the DHT problem defined as in Sections II and III, Type-I and Type-II error probabilities are bounded with the exponential forms with the following exponents}
		\begin{align}
			E_0 &= D_b\left(\frac{\gamma_t}{n}||p_0\right)\!,\\
			E_1 &= 1 - H_b\left(\frac{\gamma_t}{n}\right)\!,
		\end{align}
		with $p_0 \leq \frac{\gamma_t}{n} \leq \frac{1}{2}$. Here, $D_b\left(p||q\right)$ represents the binary Kullback-Leibler divergence between the probability pair $\left(p,q\right)$, where $\left(p,q\right) \in \left[0,1\right]$, and $H_b\left(\cdot\right)$ denotes the binary entropy function.
	\end{propstyle}
	\textit{proof}: The proof is given in Appendix B.
\begin{corstyle}
\textcolor{black}{Considering the DHT problem defined in Sections II and utilizing an $[n,k]_\rho$ as the local quantizer with a sufficiently large blocklength $n$, the \textit{only} effective parameter influencing Type-I and Type-II error probabilities is the \textit{normalized decision threshold} $\frac{\gamma_t}{n}$, when $p_0 \leq \frac{\gamma_t}{n} \leq \frac{1}{2}$.}
\end{corstyle}
\textit{proof}: \textcolor{black}{The proof directly follows from Proposition 1. \hfill $\blacksquare$
}
\section{Numerical Results}\label{section 4}
%\subsection{Optimal coset leader spectrum}
We set $\rho = n - k$ \textcolor{black}{to test Algorithm 1}, according to the Singleton bound.
\textcolor{black}{For certain values of $n$, $k$ provided in Table I}, the optimal solution precisely matches the \textit{coset leader spectrum} of \textcolor{black}{existing} linear codes\footnote{Remarkably, these codes exhibit an optimal covering radius.}.
\textcolor{black}{It should be noticed that}, however, in \textcolor{black}{some} other cases, the solutions do \textcolor{black}{\textit{not}} correspond, to \textcolor{black}{our best} knowledge, \textcolor{black}{to} the \textit{coset leader spectrum} of any known linear block code\textcolor{black}{, indicating that \textcolor{black}{optimal DHT quantizers} should not necessarily \textcolor{black}{be} based on \textcolor{black}{existing} linear codes}.%For instance, by solving the specified ILP for $ n = 31 $ and $ k = 11 $, the resulting coset distribution is as follows
\begin{figure}[t]
\centering
\begin{subfigure}{0.48\textwidth}
\centering
\includegraphics[width=\linewidth]{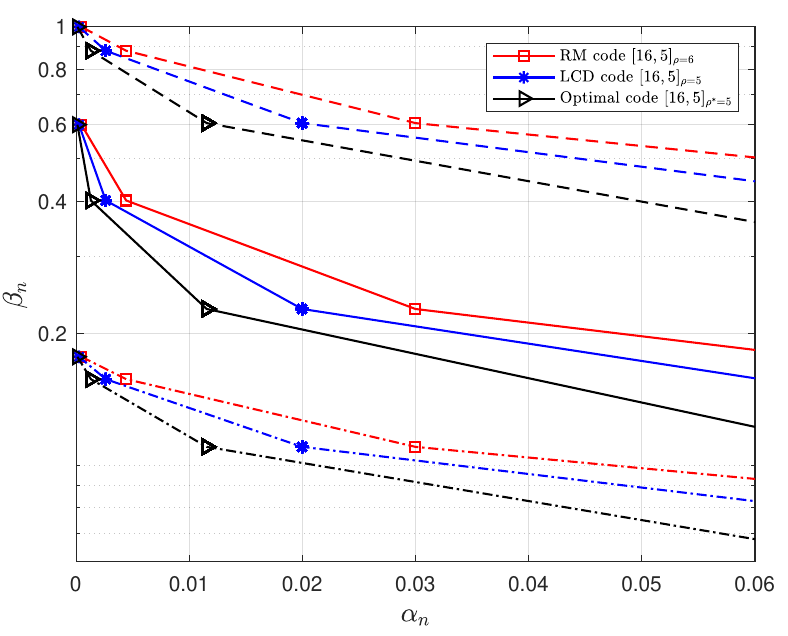}
\captionsetup{font=footnotesize, labelfont=bf}
\caption{\textcolor{black}{ROC curve for Reed-Muller (RM) code $[16,5]_{\rho=6}$, linear complementary dual (LCD) code $[16,5]_{\rho=5}$, and the optimization result code $[16,5]_{\rho^*=5}$}.}	
\label{Fig2}
\end{subfigure}
\hfill
\begin{subfigure}{0.48\textwidth}
\centering
\includegraphics[width=\linewidth]{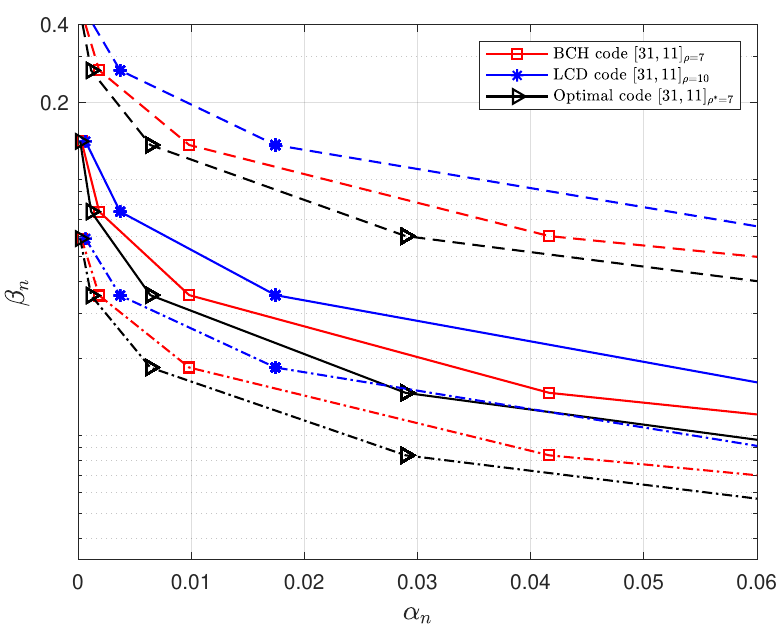}
\captionsetup{font=footnotesize, labelfont=bf}
\caption{\textcolor{black}{ROC curve for BCH code $[31,11]_{\rho=7}$, linear complementary dual (LCD) code $[31,11]_{\rho=10}$, and the optimization result code $[31,11]_{\rho^*=7}$}.}	
\label{Fig3}
\end{subfigure}
\caption{\textcolor{black}{ROC curves: exact values (solid lines), lower bounds (dash-dotted lines), and upper bounds (dashed lines)}.}
\end{figure}
\begin{figure}
\centering
\includegraphics[scale=0.7]{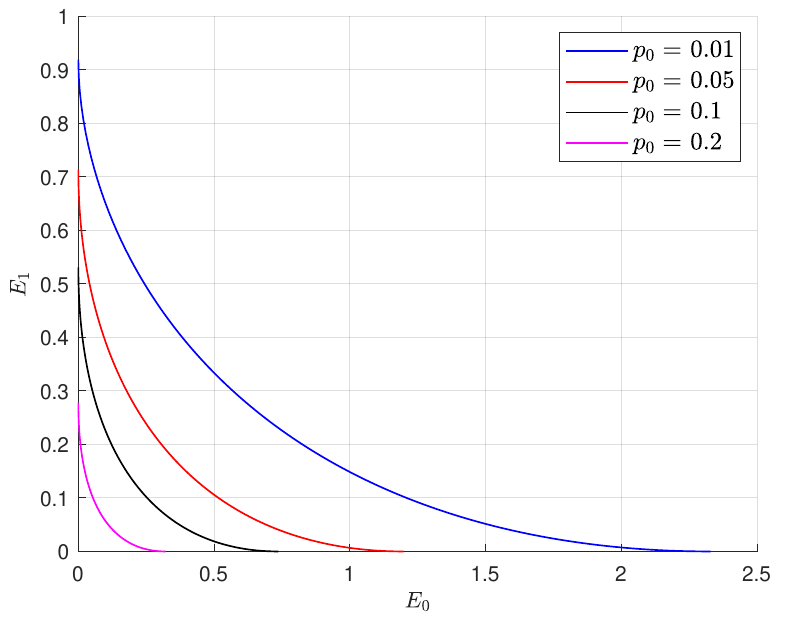}
\captionsetup{font=footnotesize, labelfont=bf}
\caption{\textcolor{black}{The tradeoff between Type-I and Type-II in terms of $E_0$ and $E_1$.}}	
\label{Fig4}
\end{figure}
We investigate Type-II error \textcolor{black}{probability, given} \textcolor{black}{the constraint $\epsilon\leq 0.06$} on Type-I error \textcolor{black}{probability and} $p_0 = 0.05$.
\textcolor{black}{Fig. \ref{Fig2} and Fig. \ref{Fig3}} \textcolor{black}{\textcolor{black}{provide} Type-I versus Type-II error probabilities, referred to Reciever Operating Characteristic (ROC) curve}, \textcolor{black}{where, the lower and upper bounds are shown using Proposition 1;}
Fig. \ref{Fig2} \textcolor{black}{considers} the Reed-Muller code $[16,5]_{\rho=6}$ \textcolor{black}{as well as the} linear complementary dual (LCD) code $[16,5]_{\rho=5}$ compared to hypothetical linear code $[16,5]_{\rho^*=5}$ obtained \textcolor{black}{by Algorithm 1}. Similarly, Fig. \ref{Fig3} presents the results for the case of the BCH code $[31,11]_{\rho=7}$ and the LCD code $[31,11]_{\rho=10}$ compared to the hypothetical linear code $[31,11]_{\rho^*=7}$ derived from \textcolor{black}{Algorithm 1}.\textcolor{black}{The numerical results demonstrate that as the covering radius $\rho$ of $[n,k]_\rho$ decreases, we approach $[n,k]_{\rho^*}$ with $\rho^*$ satisfying the equality in the Sphere-covering bound \cite[Table I]{graham1985covering}.
\textcolor{black}{It is noteworthy that the gap between the lower \textcolor{black}{and upper} bounds for $[n,k]_{\rho^*}$ and $[n,k]_\rho$ with equal $\rho$ is attributed to the difference in the decision thresholds from the AO algorithm and the thresholds chosen according to the NP criterion, respectively.}
Moreover, it is observed that as the code length n increases from $n = 16$ in Fig. \ref{Fig2} to $n = 31$ in Fig.\ref{Fig3}, the gap between the lower \textcolor{black}{and upper} bounds and the exact Type-II error probability decreases}.

\textcolor{black}{In Fig. \ref{Fig4}, the results are generated utilizing Proposition 1 for $p_0 \leq \frac{\gamma_t}{n}\leq \frac{1}{2}$. The figure indicates that the smaller $p_0$ value, i.e., the smaller value of the lower bound for $\frac{\gamma_t}{n}$, the larger exponents for Type-I and Type-II error probabilities}.   
\iffalse
\begin{align}
C^*(x)=1& + 31x + 465x^2 + 4495x^3 + 31265x^4 \nonumber\\
& + 169911x^5 + 736281x^6 + 105927x^7\label{33}.
\end{align}
The hypothetical code corresponding to (\ref{33}) has $ \rho^*=7 $. This achieves the minimum possible covering radius. The best-known linear code $ [31,11] $ has $ \rho=7 $ and its coset spectrum is \cite{bosma1997magma},
\begin{align}
C(x)=1& + 31x + 465x^2 + 4495x^3 + 31465x^4 \nonumber\\
& + 169911x^5 + 522009x^6 + 320199x^7. 
\end{align}
\fi
%In this way, by blackucing the threshold, Type-II error probability decreases and Type-I error probability increases with respect to the constraint $\epsilon$.
\section{Conclusions}\label{section 5}
In this \textcolor{black}{paper}, we have derived optimal values of Type-I and Type-II error probabilities %in DHT against independence, considering the BSS model and the NP criterion, 
through the utilization of an AO algorithm. This algorithm operates directly \textcolor{black}{based on the} analytical expressions of the error probabilities, and optimizes the \textit{coset leader spectrum} of a hypothetical binary linear code while simultaneously optimizing the decision threshold. \textcolor{black}{Numerical} results \textcolor{black}{show} the effectiveness of linear codes with optimal covering radius, approaching near-optimality.% Furthermore, we have derived lower and upper bounds for Type-I and Type-II error probabilities, \textcolor{black}{yielding} the establishment of error exponents for the large code-length regime.
%The quest for binary linear codes with optimal coset leader spectrum, in alignment with coding theory principles, remains an open question for obtaining optimal DHT codes.

\section*{Acknowledgement}
This work has received a French government support granted to the Cominlabs excellence
laboratory and managed by the National Research Agency in the Investing for the Future program
under reference ANR-10-LABX-07-01.
\begin{center}
\begin{table}[t]
\centering
\captionsetup{font=scriptsize, labelfont=bf, labelsep=period, format=plain, justification=centerlast} 
\caption{COSET LEADER SPECTRUM OBTAINED AS SOLUTIONS, IN THE INDICATED CASES, OF ALGORITHM 1}
\begin{tabular}{ |c|c|c|c|}
\hline
Optimal linear block code & $  n $ & $ k $ & $ \textcolor{black}{\mathbf{N}}^* $ \\ 
\hline
Hamming & $ 7 $ & $ 4 $ & $ \left(1,7\right) $  \\  
\hline
Reed-Muller & $ 8 $ & $ 4 $ & $ \left(1,8,7\right) $  \\
\hline  
Hamming & $ 15 $ & $ 11 $ & $ \left(1,15\right) $ \\
\hline
Golay & $ 23 $ & $ 12 $ &  $ \left(1, 23, 253, 1771\right) $\\
\hline
Extended Golay & $ 24 $ & $ 12 $ & $ \left(1, 24, 276, 2024, 1771\right) $\\
\hline
Hamming & $ 31 $ & $ 26 $ & $ \left(1, 31\right) $ \\
\hline
\end{tabular}
\end{table}
\end{center}

%\begin{figure}[t]
%	\includegraphics[scale=0.5]{coset lesder spectrum [34,15]_cropped.pdf}
%	\captionsetup{font=footnotesize, labelfont=bf}
%	\caption{Optimized \textit{coset leader} spectrum, based on integer linear program, for $ n=34 $, $ k=15 $ and $p_0=0.05 $. }	
%	\label{Fig1}
%\end{figure}

%\begin{figure}[h!]
%	\includegraphics[scale=0.5]{Type-II Improvement_cropped.pdf}
%	\captionsetup{font=footnotesize, labelfont=bf}
%	\caption{ROC comparision curves for $ n=16 $, $ k=5 $ and $ n=31 $, $ k=11 $.}.	
%	\label{Fig3}
%\end{figure}
%\begin{figure}[h]
%	\includegraphics[scale=0.5]{Threshold optimization_cropped.pdf}
%	\captionsetup{font=footnotesize, labelfont=bf}
%	\caption{Threshold optimization for $ n=16 $, $ k=5 $ and $ n=31 $, $ k=11 $.}	
%	\label{Fig4}
%\end{figure}

%\section*{Financial Information}
%This work has been conducted, in part, under the financial
%support of French National Research Agency, reference ANR-
%10-LABX-07-01.
\appendices
\section{Proof of Lemma 1}
\textcolor{black}{\textit{Proof}: According to (\ref{6}), we have
	\begin{equation}
		W_{\beta_{i}} = \left(\frac{1}{2}\right)^n\sum_{\gamma=0}^{\gamma_t}\sum_{u=0}^{\min\left(\gamma,i\right)}{i \choose u}{n-i \choose \gamma-u}.  \label{34}
	\end{equation}
	Consider the \textcolor{black}{summation} \textcolor{black}{term with respect to} $ u $ in (\ref{34}), we separate it into the following two cases:\\
	i) if \( \gamma < i \), by utilizing \textcolor{black}{the} \textcolor{black}{Vandermonde's identity \cite{comtet2012advanced}}, we have
	\begin{equation}
		\sum_{u=0}^{\gamma}{i \choose u}{n-i \choose \gamma-u} = {n \choose \gamma}. \nonumber
	\end{equation}
	ii) if \( \gamma > i \), by using the binomial formula, we have %the following set of equalities:
	\begin{small}
		\begin{align}
			\left(1+x\right)^n &= \sum_{\gamma=0}^{n} {n \choose \gamma}x^\gamma, \nonumber\\
			\left(1+x\right)^i \left(1+x\right)^{n-i} &= \sum_{u=0}^{i}{i \choose u}x^u \sum_{u^{\prime}=0}^{n-i} {n-i \choose  u^{\prime}} x^{u^{\prime}} \sum_{\gamma=0}^{n} \delta_{\gamma,u+u^{\prime}} \nonumber \\
			&=\sum_{\gamma=0}^{n}x^\gamma \sum_{u=0}^{i}{i \choose u} \sum_{u^\prime=0}^{n-i} {n-i \choose u^\prime} \delta_{u^\prime, \gamma-u}\nonumber\\
			&= \sum_{\gamma=0}^{n}x^\gamma \sum_{u=0}^{i}{i \choose u}{n-i \choose \gamma-u} \Big|_{0 \leq \gamma-u \leq n-i},\nonumber
		\end{align}
	\end{small}
	\textcolor{black}{where $\delta_{\left(\cdot,\cdot\right)}$ denotes the Kronecker \textcolor{black}{delta} function}. \textcolor{black}{Then, the following equality is derived:} \vspace{-0.1in}
	\begin{equation}
		\sum_{\substack{u=0\\ \gamma-\left(n-i\right) \leq u \leq \gamma}}^{i} {i \choose u}{n-i \choose \gamma-u} = {n \choose \gamma},\label{18}
	\end{equation}
	given that $ \gamma-\left(n-i\right) =  \gamma-n <0 $ when $ i = 0 $ is the minimum possible value for the summation index, the summation in (\ref{18}) is properly bounded. Therefore, we can express (\ref{25}) as,
	\begin{align}
		\beta_n&=\frac{1}{N}\left(\frac{1}{2}\right)^n\sum_{i=0}^{\textcolor{black}{\rho}} \sum_{\gamma=0}^{\gamma_t} {n \choose \gamma}N_{i}\nonumber\\
		&= \left(\frac{1}{2}\right)^n\sum_{\gamma=0}^{\gamma_t}{n \choose \gamma}.\nonumber \tag*{\hfill$\blacksquare$}
	\end{align}
}
\section{Proof of Proposition 1}
In this appendix, we provide a more detailed \textit{proof} of Proposition 1. Based on Eq.~(\ref{7}), Type-I error probability is given by:
\begin{align}
	\alpha_n & =  \frac{1}{N} \sum_{\gamma=\gamma_t+1}^{n}\sum_{i=0}^{\rho}\sum_{u=0}^{\min\left(\gamma,i\right)} {i \choose u}{n-i \choose \gamma-u} p_0^{\left(\gamma+i-2u\right)} (1-p_0)^{n-(\gamma+i-2u)}\\
	&\overset{(a)}{\leq} \frac{1}{N} \sum_{\gamma=\gamma_t+1}^{n}\sum_{i=0}^{\rho}\sum_{u=0}^{\min\left(\gamma,i\right)} {n \choose \gamma} p_0^{\left(\gamma+i-2u\right)}\left(1-p_0\right)^{n-\left(\gamma+i-2u\right)}N_i\\
	&=\frac{1}{N} \sum_{\gamma=\gamma_t+1}^{n} {n \choose \gamma} p_0^\gamma \left(1-p_0\right)^{n-\gamma} \sum_{i=0}^{\rho}  \left(\frac{p_0}{1-p_0}\right)^i \sum_{u=0}^{\min\left(\gamma,i\right)} \left(\frac{1-p_0}{p_0}\right)^{2u}N_i\\
	& \overset{(b)}{=} \frac{1}{N} \sum_{\gamma=\gamma_t+1}^{n} {n \choose \gamma} p_0^\gamma \left(1-p_0\right)^{n-\gamma} \underbrace{\sum_{i=0}^{\rho} \left(\frac{p_0}{1-p_0}\right)^i \left[\frac{\left(\left(\frac{1-p_0}{p_0}\right)^2\right)^{\min\left(\gamma,i\right)}-1}{\left(\frac{1-p_0}{p_0}\right)^2-1}\right]}_{B}N_i ,\label{upper bound}
\end{align}	
Thus, the inequality $\left(a\right)$ is obtained using Vandermonde's identity \cite{comtet2012advanced} as ${i \choose u}{n-i \choose \gamma-u} \leq {n \choose \gamma}$, and the equality $\left(b\right)$ is derived from the formula for the geometric sum with the ratio $\kappa = \left(\frac{1-p_0}{p_0}\right)^2$. On the other hand, we can write:
\begin{align}
	&B = \sum_{i=0}^{\rho} \left(\frac{1}{\kappa}\right)^{\frac{i}{2}} \left[\frac{\kappa^{\min\left(\gamma,i\right)}-1}{\kappa-1}\right]N_i \nonumber\\
	& \overset{(c)}{\leq} \sum_{i=0}^{\rho} \left(\frac{1}{\kappa}\right)^{\frac{i}{2}} \textcolor{black}{\left[\frac{\kappa^{i}-1}{\kappa-1}\right]}N_i \nonumber\\
	%& \leq \sum_{i=0}^{\rho} \left(\frac{1}{\kappa}\right)^{\frac{i}{2}} \textcolor{black}{\left[\kappa^{i}-1\right]}N_i  \nonumber
	%& \triangleq \Delta, \label{B}
	&\leq \sum_{i=0}^{\rho} \left(\frac{1}{\kappa}\right)^{\frac{i}{2}} \textcolor{black}{\left[\kappa^{i}-1\right]}N_i  \nonumber\\
	&\triangleq \Delta^\prime, \label{B}
\end{align}
where, $\left(c\right)$ holds because $\kappa \geq 1$ and also by the property of the minimum function, $\min(\gamma,i) \leq i$. Therefore, by substituting Eq. (\ref{B}) into Eq. (\ref{upper bound}), we have:
\begin{align}
	\alpha_n & \leq \frac{\Delta^\prime}{N} \sum_{\gamma=\gamma_t+1}^{n} {n \choose \gamma} p_0^\gamma \left(1-p_0\right)^{n-\gamma}\nonumber\\
	& \overset{(d)}{\leq} \frac{\Delta^\prime}{N} 2^{-nD_b\left(\frac{\gamma_t+1}{n}||p_0\right)},
\end{align}
If \(p_0 \leq \frac{\gamma_t + 1}{n} \leq 1\), the inequality $\left(d\right)$ follows from the upper bound related to the binomial distribution in \cite[lemma 4.7.2]{ash2012information}.

Next, we will calculate the lower bound related to Type-I error probability. To do this, consider Stirling's approximation as follows \cite{prugel2020probability}
\begin{align}
	\sqrt{2\pi}n^{\left(n+\frac{1}{2}\right)}e^{-n} \leq n! \leq n^{\left(n+\frac{1}{2}\right)}e^{1-n}.\label{Stirling-bound}
\end{align}
Thus, we can obtain the lower bound related to the binomial coefficients below using (\ref{Stirling-bound})
\begin{align}
	{i \choose u} &\geq \frac{\sqrt{2\pi}e^{-i}i^{\left(i+\frac{1}{2}\right)}}{e^{1-u}u^{\left(u+\frac{1}{2}\right)}e^{1-\left(i-u\right)}\left(i-u\right)^{\left(i-u+\frac{1}{2}\right)}} \nonumber\\
	&= \sqrt{2\pi}e^{-2}\sqrt{\frac{i}{u\left(i-u\right)}} 2^{iH_b\left(\frac{u}{i}\right)}, \label{lowercom1}
\end{align}
\begin{align}
	{n-i \choose \gamma-u} &\geq \sqrt{2\pi} e^{-2} \sqrt{\frac{\left(n-i\right)}{\left(\gamma-u\right)\left(n-i-\gamma+u\right)}} 2^{\left(n-i\right)H_b\left(\frac{\gamma-u}{n-i}\right)}, \label{lowercom2}
\end{align}   
and also
\begin{equation}
	{n \choose \gamma} \leq \frac{e}{2\pi} \sqrt{\frac{n}{\gamma(n - \gamma)}} 2^{n H_b\left(\frac{\gamma}{n}\right)}.\label{uppercom1}
\end{equation}                                            
On the other hand, since the Type-I error probability can be expressed as follows
\begin{align}
	\alpha_n = \frac{1}{N} \sum_{\gamma=\gamma_t+1}^{n} {n \choose \gamma} p_0^\gamma \left(1-p_0\right)^{n-\gamma} \sum_{w=0}^{\rho} \left(\frac{p_0}{1-p_0}\right)^{i} \sum_{u=0}^{\min\left(\gamma,i\right)} \left(\frac{1-p_0}{p_0}\right)^{2u} \frac{{i \choose u}{n-i \choose \gamma-u}}{{n \choose \gamma}}N_i,\nonumber
\end{align}
Therefore, using (\ref{lowercom1}), (\ref{lowercom2}), and (\ref{uppercom1}), we will have
\begin{align}
	\alpha_n &\geq \frac{1}{N} \sum_{\gamma=\gamma_t+1}^{n} {n \choose \gamma} p_0^\gamma \left(1-p_0\right)^{n-\gamma} \sum_{i=0}^{\rho}  \left(\frac{p_0}{1-p_0}\right)^i \sum_{u=0}^{\min\left(\gamma,i\right)} \left(\frac{1-p_0}{p_0}\right)^{2u}N_i\nonumber\\
	& \times (4\pi^2) e^{-3} \sqrt{\frac{i \gamma \left(n-\gamma\right) \left(n-i\right)}{u \left(i-u\right) n \left(\gamma-u\right) \left(n-i-\gamma+u\right)}} 2^{\left[iH_b\left(\frac{u}{i}\right) + \left(n-i\right)H_b\left(\frac{\gamma-u}{n-i}\right) - nH_b\left(\frac{\gamma}{n}\right)\right]}\nonumber\\
	& \overset{(e)}{\geq} \frac{1}{N} \sum_{\gamma=\gamma_t+1}^{n} {n \choose \gamma} p_0^\gamma \left(1-p_0\right)^{n-\gamma} \sum_{i=0}^{\rho} \left(\frac{p_0}{1-p_0}\right)^i \sum_{u=0}^{\min\left(\gamma,i\right)} \left(\frac{1-p_0}{p_0}\right)^{2u} N_i \nonumber\\
	&\times \left(4\pi^2\right) e^{-3} \sqrt{\frac{\gamma}{n.i}} 2^{\left[iH_b\left(\frac{u}{i}\right) + \left(n-i\right)H_b\left(\frac{\gamma-u}{n-i}\right) - nH_b\left(\frac{\gamma}{n}\right)\right]}\nonumber\\
	& \overset{(f)}{\geq} \frac{1}{N} \sum_{\gamma=\gamma_t+1}^{n} {n \choose \gamma} p_0^\gamma \left(1-p_0\right)^{n-\gamma} \sum_{i=0}^{\rho} \left(\frac{p_0}{1-p_0}\right)^i \sum_{u=0}^{\min\left(\gamma,i\right)} \left(\frac{1-p_0}{p_0}\right)^{2u} N_i \nonumber\\
	&\times \left(4\pi^2\right)e^{-3} \sqrt{\frac{\gamma}{n.i}} 2^{-nH_b(\frac{\gamma}{n})}\nonumber\\
	& \overset{(g)}{\geq} \frac{\left(4\pi^2\right) e^{-3}}{N} \sum_{\gamma=\gamma_t+1}^{n} {n \choose \gamma} p_0^\gamma \left(1-p_0\right)^{n-\gamma} \sum_{i=1}^{\rho} (\frac{1}{i})^{\frac{1}{2}}\left(\frac{p_0}{1-p_0}\right)^i \sum_{u=0}^{\min\left(\gamma,i\right)} \left(\frac{1-p_0}{p_0}\right)^{2u} N_i \nonumber\\
	& \overset{(h)}{=} \frac{\left(4\pi^2\right)e^{-3}}{N}   \sum_{\gamma=\gamma_t+1}^{n} {n \choose \gamma} p_0^\gamma \left(1-p_0\right)^{n-\gamma} \sum_{i=0}^{\rho} (\frac{1}{i})^{\frac{1}{2}}\left(\frac{p_0}{1-p_0}\right)^i \left[\frac{ \left(\frac{1-p_0}{p_0}\right)^{2\min\left(\gamma,i\right)} - 1}{\left(\frac{1-p_0}{p_0}\right)^2 - 1}\right]N_i \nonumber\\
	& \overset{(n)}{\geq} \frac{\left(4\pi^2\right)e^{-3}}{N}  \sum_{\gamma=\gamma_t+1}^{n} {n \choose \gamma} p_0^\gamma \left(1-p_0\right)^{n-\gamma} \sum_{i=0}^{\rho} (\frac{1}{i})^{\frac{1}{2}}\left(\frac{p_0}{1-p_0}\right)^i \left[\frac{ 1}{\left(\frac{1-p_0}{p_0}\right)^2}\right]N_i\nonumber\\
	& = \frac{\left(4\pi^2\right)e^{-3}}{N}  \sum_{\gamma=\gamma_t+1}^{n} {n \choose \gamma} p_0^\gamma \left(1-p_0\right)^{n-\gamma} \sum_{i=0}^{\rho} (\frac{1}{i})^{\frac{1}{2}} \left(\frac{p_0}{1-p_0}\right)^{\left(i+\textcolor{black}{2}\right)}N_i \nonumber\\
	&\overset{(m)}{\geq} \frac{\Delta^\prime
	}{N}  \left[8\left(\gamma_t + 1\right)\left(1 - \frac{\gamma_t + 1}{n}\right)\right]^{-\frac{1}{2}} 2^{-nD_b\left(\frac{\gamma_t + 1}{n} \middle\| p_0\right)},
\end{align}
So that \(\left(e\right)\) arises from the fact that \(\gamma \leq n\), \(u \leq i\), and \((\gamma-u) \leq (n-i)\). Additionally, \(\left(f\right)\) is derived from the concavity of the function \(H_b\left(\cdot\right)\) using Jensen's Inequality \cite{thomas2006elements}. On the other hand, \(\left(g\right)\) is obtained based on numerical results by choosing \(\gamma_t=n\) within the range \(\gamma_t+1 \leq \gamma \leq n\), which achieves the maximum value of the coefficient \(\sqrt{\frac{\gamma}{n}} 2^{-nH_b(\frac{\gamma}{n})}\) and thus leads to a tighter lower bound. The equality \(\left(h\right)\) is derived using the sum of a geometric series with a ratio of \(\kappa = \left(\frac{1-p_0}{p_0}\right)^2\). Furthermore, \(\left(n\right)\) holds because in the interval \(0 \leq p_0 < \frac{1}{2}\), we have \(\kappa \geq 0\). Finally \(\left(m\right)\) is defined as \(\Delta^\prime = (4\pi^2) e^{-3} \sum_{i=0}^{\rho} \left(\frac{1}{i}\right)^{\frac{1}{2}} \left(\frac{p_0}{1-p_0}\right)^{(i+2)} N_i\) and is obtained using the lower bound related to the binomial distribution in \cite[Lem 4.7.2]{ash2012information} within the allowable range \(p_0 \leq \frac{\gamma_t + 1}{n} \leq 1\).

According to Lemma 1, we have:

\begin{equation}
	\beta_n = \left(\frac{1}{2}\right)^n \sum_{\gamma=0}^{\gamma_t} {n \choose \gamma}.\nonumber
\end{equation}
By a similar argument, when \(0 \leq \frac{\gamma_t}{n} \leq \frac{1}{2}\), we can write:
\begin{align}
	\left[8\gamma_t\left(1 - \frac{\gamma_t}{n}\right)\right]^{-\frac{1}{2}} 2^{-nD_b\left(1-\frac{\gamma_t}{n}||\frac{1}{2}\right)} \leq \beta_n \leq 2^{-nD_b\left(1-\frac{\gamma_t}{n}||\frac{1}{2}\right)}. 
\end{align}
\textcolor{black}{such that for sufficiently $n$} %the lower bound of $\alpha_n$ depends on the covering radius $\rho$ and the normalized threshold $\frac{\gamma_t}{n}$, also the lower bound of $\beta_n$ only depends on $\frac{\gamma_t}{n}$.}
\begin{small}
\begin{align}
	E_0 &  =- \left\{\frac{1}{n} \log \alpha_n\right\}  \qquad ; \quad n \rightarrow \text{large} \nonumber \\
	&  = D_b\left(\frac{\gamma_t}{n}||p_0\right) \;\;\; \qquad ; \quad p_0 \leq \frac{\gamma_t}{n} \leq 1 \nonumber
\end{align}
\end{small}
and 
\begin{small}
\begin{align}
	E_1	&  = -\left\{ \frac{1}{n} \log \beta_n\right\}   \qquad ; \quad n \rightarrow \text{large}\nonumber\\ 
	&  = D_b\left(1-\frac{\gamma_t}{n}||\frac{1}{2}\right)\nonumber\\
	& \overset{(f)}{=} 1- H_b\left(\frac{\gamma_t}{n}\right) \; \qquad ; \quad 0 \leq \frac{\gamma_t}{n} \leq \frac{1}{2} \nonumber
\end{align}
\end{small}where $(f)$ arises from the symmetry property of the binary entropy function $ H_b\left(\cdot\right) $, when $\frac{\gamma_t}{n} \leq \frac{1}{2} $. \hfill $\blacksquare$

\newpage
\bibliographystyle{ieeetran}
\bibliography{isit}

\end{document}